\documentclass[11pt]{article}

\usepackage{geometry}
\usepackage{fullpage}
\usepackage{graphicx}
\usepackage{pdflscape}
\usepackage{multirow,multicol}
\usepackage{ragged2e}

\usepackage{dcolumn}
\usepackage{booktabs,calc}
\usepackage{longtable}
\usepackage{array}
\usepackage{epigraph}	
\usepackage{paralist}
\usepackage{verbatim}
\usepackage{subfig}
\usepackage{setspace}
\usepackage{amsmath}
\usepackage{amssymb}
\usepackage{natbib}
\usepackage[linktocpage=true, colorlinks=true, linkcolor=blue, citecolor=black]{hyperref}
\usepackage{enumerate}
\usepackage{authblk}
\usepackage{placeins}

\DeclareMathOperator{\E}{\mathbb{E}}
\DeclareMathOperator{\V}{\mathbb{V}}

\DeclareMathOperator*{\argmin}{arg\,min}

\begin{document}
	
\title{Recent Developments in Inference: Practicalities for Applied Economics\thanks{Correspondence to be sent to: jdmichler@arizona.edu. The authors gratefully acknowledge the helpful comments from Laura Bakkensen, Leah Bevis, Jeffrey Bloem, Metin \c{C}akır, Michael Delgado, Jill Hobbs, Jason Kerwin, Connor Mullally, A. Ford Ramsey, Jacob Ricker-Gilbert, Jutta Roosen, and David Ubilava.}}
	
	\author[a]{Jeffrey D. Michler}
	\author[a]{Anna Josephson}
	\affil[a]{\small \emph{Department of Agricultural and Resource Economics, University of Arizona}}

\date{July 2021}
\maketitle
	
\begin{center}\begin{abstract}
	\noindent We provide a review of recent developments in the calculation of standard errors and test statistics for statistical inference. While much of the focus of the last two decades in economics has been on generating unbiased coefficients, recent years has seen a variety of advancements in correcting for non-standard standard errors. We synthesize these recent advances in addressing challenges to conventional inference, like heteroskedasticity, clustering, serial correlation, and testing multiple hypotheses. We also discuss recent advancements in numerical methods, such as the bootstrap, wild bootstrap, and randomization inference. We make three specific recommendations. First, applied economists need to clearly articulate the challenges to statistical inference that are present in data as well as the source of those challenges. Second, modern computing power and statistical software means that applied economists have no excuse for not correctly calculating their standard errors and test statistics. Third, because complicated sampling strategies and research designs make it difficult to work out the correct formula for standard errors and test statistics, we believe that in the applied economics profession it should become standard practice to rely on asymptotic refinements to the distribution of an estimator or test statistic via bootstrapping. Throughout, we reference built-in and user-written Stata commands that allow one to quickly calculate accurate standard errors and relevant test statistics.
\end{abstract}\end{center}

{\small \noindent\emph{JEL Classification}: A23, B41, C12, C15, C87
}


\newpage	
\doublespacing
\section{Introduction}
The last two decades have seen phenomenal growth in applied microeconomic research. This growth stems from the causal revolution and the associated econometric methods to control for endogenous regressors \citep{angrist_mostly_2009}. Many of the advances are framed within a program evaluation approach, which relies on simpler, reduced form econometric techniques, with a heavy dependence on OLS \citep{abadie_econometric_2018}. While the program evaluation approach to policy analysis is not without its critics \citep{deaton_instruments_2010, heckman_building_2010}, it has permeated all fields of economics. Applied economists use experimental methods (randomized control trials in the lab or field) or quasi-experimental methods (regression discontinuity, difference-in-difference, or instrumental variables) to estimate treatment effects from an intervention. An ever-expanding literature exists on obtaining consistent and unbiased estimates of these treatment effects \citep{deaton_instruments_2010, heckman_building_2010, imbens_better_2010, belloni_high-dimensional_2014, imbens_causal_2015, morgan_counterfactuals_2015, abadie_econometric_2018}.

In this paper, we address an altogether different problem with causal inference within a program evaluation context. We assume away biases in estimating causal effects and instead focus on the issues relating to the standard error of the estimand. Standard errors are used to make inference on the statistical significance of causal parameters. Yet, how to correctly calculate standard errors, and the test statistics and confidence intervals based on them, receives less attention in the applied economics literature than endogeneity-induced bias in coefficient estimates. While the issues of heteroskedasticity and serial correlation are dealt with in any econometric textbook, there are numerous challenges to inference that are not commonly covered. These include the issue of grouped data, multiple hypothesis testing, and numerical methods, such as bootstrapping and randomization inference. These topics have been addressed by a number of recent and important advancements in inference, which have even provided improvements on classic topics like heteroskedasticity and serial correlation. The goal of this paper is to synthesize these recent advancements and to provide guidance in their application.

Among applied economists today, calculating the correct standard errors and test statistics is often viewed as a second-order concern, while generating consistent and unbiased estimators is of first-order importance. This hierarchy is a natural response to the criticism by \cite{ziliak_cult_2008}, that applied economics has become overly focused on arbitrary cut-offs like $p > 0.95$, forming what they term the ``cult of statistical significance''. We agree that too much emphasis can be placed on arbitrary cut-offs and not enough emphasis on economic significance, leading to well-known issues of $p$-hacking and publication bias \citep{brodeur_methods_2020}. Yet we believe that getting the correct standard errors and test statistics is just as important as generating unbiased estimates of coefficients.

To illustrate this, imagine four scenarios applied economists might find themselves in: (1) an unbiased estimate with incorrect (too small) standard errors that results in rejection of the null; (2) an unbiased estimate with incorrect (too large) standard errors that result in failure to reject the null; (3) a biased estimate with correct standard errors that result in rejection of the null; and (4) a biased estimate with correct standard errors that results in failure to reject the null. If we adopt Angrist and Pischke's (2009) heuristic that committing Type II errors (failing to reject a false null) are less costly than committing Type I errors (rejecting a true null), then applied economists should prefer scenarios (2) and (4). In scenario (1), one is committing a Type I error. Even though the coefficient estimate is unbiased, the standard errors are wrong, leading one to reject the true null and claim that there really is a statistically significant effect. In scenario (2), one is committing a Type II error; the coefficient estimate is unbiased but one cannot draw any conclusions as the standard errors lead one to conclude that the effect size is not statistically significant. In scenario (3), one is only potentially committing a Type I error. As the coefficient estimate is biased, and assuming the size of the bias is unknown, one does not know if the null is true or not, and so one is uncertain if they are correctly rejecting the null or making a Type I error. In scenario (4), one is potentially committing a Type II error. Here, the bias in the coefficient estimate means one is uncertain about the true effect size and so is unsure if the failure to reject the null is accurate or a Type II error.

Obviously, the preferred scenario is unbiased estimates and correct standard errors. But, the point is to emphasize that the hierarchy of bias over precision may lead applied economists astray. In scenario (1), one would mistakenly draw conclusions about the importance of the policy under consideration, even though the problem of endogeneity has been addressed and the unbiased coefficient estimate has been calculated. In scenario (4), one cannot be certain if an error is being committed but can be certain one is not committing a Type I error and incorrectly claiming a statistically significant causal effect when there is none. \cite{vivalt_how_2021} term this de-emphasis on statistical inference ``variance neglect'', in which individuals pay less attention to the variance of the estimand than a Bayesian updater would. Again, we are not arguing that precision is more important than bias or that applied economists and policymakers should focus on arbitrary cut-offs, like $p > 0.95$. Rather, we believe that both are equally important for applied economic analysis.

We make three specific points. First, applied economists need to clearly articulate the challenges to statistical inference that are present in data as well as the source of those challenges. As \cite{abadie_samplingbased_2020} argue, there is much muddled thinking about inference in the literature and many of the heuristic approaches to the calculation of standard errors and test statistics can be improved upon. To facilitate applied economists in better articulating how they arrived at the correct standard errors, we present the challenges to inference and the available solutions in an intuitive and informal way. We do not provide a mathematically detailed or rigorous treatment of either asymptotic theory or asymptotic refinements. Our goal is to avoid much of the jargon of econometric textbooks and provide applied economists with the intuition to understand issues in getting accurate inference. Second, modern computing power and statistical software means that applied economists have no excuse for not correctly calculating their standard errors and test statistics. User-friendly code for implementation almost always exists, even for the most recent advancements in statistical inference. To make the chapter as practical as possible, we provide reference to built-in and user-written Stata commands that allow one to quickly calculate accurate standard errors and relevant test statistics. Third, complicated sampling strategies and research designs make it difficult to work out the correct formula for the standard error around a specific estimand. In these cases, numerical resampling (bootstrapping) can provide approximations to the distribution of an estimator or test statistics as good as or better than those provided by asymptotic theory. Because of this, we believe that in the applied economics profession it should become standard practice to rely on asymptotic refinements to the distribution of an estimator or test statistic, presenting appropriately calculated bootstrapped critical values for hypothesis testing in the main results, and not hypothesis tests that rely on first-order asymptotic theory.

The remainder of this paper is organized as follows: we begin with a review of the problem of statistical inference and discuss the difference between a sample-based and design-based approach to inference. The next section lays out the challenges to classical inference, focusing on the problems of heteroskedasticity, grouped data, serial correlation, and multiple hypothesis testing. The penultimate section discusses the value of asymptotic refinements, including classical bootstrapping, residual (wild) bootstrapping, and randomization inference. The final section concludes with our recommendations for applied economists. 


\section{Review of the Problem}

To fix ideas, we briefly outline the problem of inference using the textbook sampling-based approach \citep{cameron_microeconometrics_2005, angrist_mostly_2009, wooldridge_econometric_2010} followed by the design-based approach \citep{abadie_when_2017, abadie_samplingbased_2020}. While throughout we have tried to minimize reliance on complicated mathematical formulas and econometric jargon, some math is necessary to clarify critical points and draw distinctions between different corrections for standard errors and test statistics. Where math and jargon are necessary, we have worked to provide intuitive explanations and definitions so as to allow those unfamiliar with matrix algebra to still grasp the key concepts.

In the textbook sampling-based approach to inference, applied economists work with a finite sample of $n$ units characterized by outcome $y_i$ and at least one $x_i \in X_i$. In this setting, uncertainty in the estimand results from the limitation that one can only observe a subset of the entire population. That one does not observe the entire population is why there is variance to the estimand. Thus, correct inference requires that applied economists know something about the sampling process. Assume that whether or not a unit from the population is in the sample is binary. We can represent whether or not the unit is sampled as $R_i \in {0,1}$. Table~\ref{tab:sample} provides an example of two different samples that one might end up with. An applied economist might randomly sample grocery shoppers at a store and survey them about their purchases, ending up with the actual sample in Table~\ref{tab:sample}. But, depending on when the sampling takes place, one might have ended up with alternative sample 1, which contains different shoppers than in the actual sample. The estimand is then a function of the observed data, $\{R_i, y_i R_i, x_i R_i \}$. One uses the observed sample to make inference about the population, but one could have just as easily ended up with an alternative sample made up of different units.

\begin{table}[htbp]	\centering
    \caption{Sampling-Based Uncertainty (\checkmark is observed, ? is missing)}  \label{tab:sample}
	\scalebox{1}
	{ \setlength{\linewidth}{.1cm}\newcommand{\contents}
		{\begin{tabular}{lcccccccc}
            \\[-1.8ex]\hline 
			\hline \\[-1.8ex]
            & \multicolumn{3}{c}{Actual Sample} &  & \multicolumn{3}{c}{Alternative Sample 1} & $\cdots$  \\
            \cline{2-4} \cline{6-8}
            Unit & $y_i$ & $x_i$ & $R_i$ & & $y_i$ & $x_i$ & $R_i$ & $\cdots$ \\
            \midrule
            1 & \checkmark & \checkmark & 1 & & \checkmark & \checkmark & 1 & $\cdots$ \\
            2 & ? & ? & 0 & & \checkmark & \checkmark & 1 & $\cdots$ \\
            3 & \checkmark & \checkmark & 1 & & ? & ? & 0 & $\cdots$ \\
            4 & ? & ? & 0 & & \checkmark & \checkmark & 1 & $\cdots$ \\
            $\vdots$ & $\vdots$ & $\vdots$ & $\vdots$ & & $\vdots$ & $\vdots$ & $\vdots$ & $\vdots$ \\
            $n$ & \checkmark & \checkmark & 1 & & ? & ? & 0 & $\cdots$ \\
			\\[-1.8ex]\hline 
			\hline \\[-1.8ex]
    		\multicolumn{9}{l}{\footnotesize \textit{Note}: Table is adapted from \cite{abadie_samplingbased_2020}.}
    	\end{tabular}}
	\setbox0=\hbox{\contents}
    \setlength{\linewidth}{\wd0-2\tabcolsep-.25em}
    \contents}
\end{table}

Our setup of the sampling-based approach allows one to make inference and conduct complex statistical tests with multiple controls, though we focus inference on the significance of a single ``treatment'' (defined $t$) variable. For a given outcome, $y$, define the conditional expectation function (CEF) as:

\begin{equation}
    \E \left[ y_i | X_i = x \right] = \int t f_y \left( t | X_i = x \right) dt.
\end{equation}

\noindent Here the CEF is the expectation, or sample mean, of individual outcomes $y_i$, given a $k \times 1$ vector of covariates $X_i$, which is held fixed. The conditional density function of $y_i$ is $f_y (t|X_i = x)$. The CEF allows us to make unbiased inference about the impact of a specific $x_k$ (a single covariate element in $X$) on the outcome for the population. Stated another way, we can use the sample CEF to learn about the population CEF.

Using the law of iterated expectations, we can rewrite the CEF as:

\begin{equation}
    y_i = \E \left[ y_i | X_i \right] + \epsilon,
\end{equation}

\noindent where $\epsilon_i$ is mean independent of $X_i$. For an arbitrary function $\beta(X_i)$, the CEF solves:

\begin{equation}
    \E \left[ y_i | X_i \right] = \argmin_{\beta(X_i)} \E \left[ \left( y_i - \beta (X_i ) \right)^2 \right],
\end{equation}

\noindent which is the minimum mean squared error prediction problem. If we consider the case where $\beta (X_i ) = \beta^{'} X_i$, in which $\beta$ is a $k \times 1$ coefficient vector, then, using the first order condition, the solution is $\hat{\beta} = \E [X_{i}^{'} X_{i}]^{-1} \E [X_{i}^{'} y_{i}]$, which is the well-known least squares estimator. Throughout, we will maintain the assumption of strict exogeneity so that estimates on all covariates can be interpreted as causal effects, which is to say: the error from the structural conditional mean contains no common causes of $y_i$ and $X_i$. The question then is, how do we ensure correct inference with respect to hypotheses about the population parameters?

To calculate test statistics, we must calculate the distribution of:

\begin{equation}
    \hat{\beta} = \left[ \sum_{i} X_{i}^{'} X_{i} \right]^{-1} \sum_{i} X_{i}^{'} y_{i}.
\end{equation}

\noindent The sampling-based approach does this by drawing repeated random samples from the population and relying on the law of large numbers for the sample estimates to get arbitrarily close to the population parameters. If, in addition to the exogeneity assumption, we also assume homoskedasticity, then $\E [ \epsilon_{i}^{2} | X_i ] = \Omega$, which is a constant. These assumptions about the first two moments of the distribution of $\epsilon$ allows us to calculate the variance of the ordinary least squares (OLS) estimator:

\begin{equation}
    \V (\beta) = \left( X_{i}^{'} X_{i} \right)^{-1} \left( X_{i}^{'} \Omega X_{i} \right) \left( X_{i}^{'} X_{i} \right)^{-1}.
\end{equation}

\noindent If we additionally assume that $\Omega$ is diagonal, meaning the sample is independent and identically distributed (\emph{i.i.d.}), then the variance reduces to the standard OLS variance:

\begin{equation}
    \V_{OLS} = \sigma^2 \left( X_{i}^{'} X_{i} \right)^{-1}
\end{equation}

\noindent where $\sigma= \Omega_{ii} = \V (\epsilon_i)$. In practice, the OLS variance is estimated using the estimated regression residual, $\hat{\epsilon}_i = y_i - \hat{\beta}^{'} X_i$, where:

\begin{equation}
    \hat{\sigma}^{2} = \sum \frac{\hat{\epsilon}^{2}_{i}}{N}
\end{equation}

\noindent is the estimated variance of the residual and $N$ is the sample size. This gives us an estimated OLS variance of $\hat{\V}_{OLS} = \hat{\sigma}^{2} (X_{i}^{'} X_{i})^{-1}$. But this estimated variance has a well-known problem: it is biased in finite samples, tending to be too small. There is an easy correction for the bias. One simply needs to correct for the degrees of freedom $(k)$ by dividing by $(N-k)$ instead of just $N$. This correction is made by default in most statistical software.

An alternative to the above sampling-based approach to uncertainty and inference is a design-based approach. Even though the sampling-based approach is now the default way that inference is presented in econometric textbooks, the design-based approach has a long historical precedent, starting with \cite{neyman_application_1923} and \cite{fisher_design_1935}.  Recently in economics, this approach has re-emerged because it provides a broader framework for thinking about different sample and population sizes, as well as different potential outcomes \citep{imbens_causal_2015, abadie_when_2017, abadie_samplingbased_2020}.

In design-based inference, uncertainty arises from an altogether different source. Imagine that for each unit of the population, there are two potential outcomes: $y_i^* (0)$ or $y_i^* (1)$. Each unit has the potential to be in one state or the other, but not both. The variable $x_i \in \{ 0,1 \}$ records which potential outcome is observed. Table~\ref{tab:design} provides an example of two different samples that one might observe. An applied economist might obtain scanner data for a grocery store that provides purchase information for the entire population of shoppers in the store revealing outcomes represented by the actual sample in Table~\ref{tab:design}. But, depending on the process that led shoppers to purchase $\left( y_i^* (1) \right)$ or not purchase $\left( y_i^* (0) \right)$ an item, one might have ended up with alternative sample 1, which contains the same shoppers as the actual sample but different outcomes. In this set-up, the estimand is a function of the observed data, $\{y_i^* (0), y_i^* (1), x_i \}$. Correct inference requires that applied economists know something about the assignment process. In design-based inference, one can observe the entire population, such as all purchases by shoppers in a grocery store, and still have uncertainty in the estimate, because of the randomness in the assignment process. Applied economists use the population observed in one state to make inference about the population in an alternative, counterfactual state.

\begin{table}[htbp]	\centering
    \caption{Design-Based Uncertainty (\checkmark is observed, ? is missing)}  \label{tab:design}
	\scalebox{1}
	{ \setlength{\linewidth}{.1cm}\newcommand{\contents}
		{\begin{tabular}{lcccccccc}
            \\[-1.8ex]\hline 
			\hline \\[-1.8ex]
            & \multicolumn{3}{c}{Actual Sample} &  & \multicolumn{3}{c}{Alternative Sample 1} & $\cdots$  \\
            \cline{2-4} \cline{6-8}
            Unit & $y_i^* (1) $ & $y_i^* (0)$ & $x_i$ & & $y_i^* (1) $ & $y_i^* (0)$ & $x_i$ & $\cdots$ \\
            \midrule
            1 & \checkmark & ? & 1 & & \checkmark & ? & 1 & $\cdots$ \\
            2 & ? & \checkmark & 0 & & ? & \checkmark & 0 & $\cdots$ \\
            3 & \checkmark & ? & 1 & & ? & \checkmark & 0 & $\cdots$ \\
            4 & \checkmark & ? & 1 & & ? & \checkmark & 0 & $\cdots$ \\
            $\vdots$ & $\vdots$ & $\vdots$ & $\vdots$ & & $\vdots$ & $\vdots$ & $\vdots$ & $\vdots$ \\
            $n$ & \checkmark & ? & 1 & & \checkmark & ? & 1 & $\cdots$ \\
			\\[-1.8ex]\hline 
			\hline \\[-1.8ex]
    		\multicolumn{9}{l}{\footnotesize \textit{Note}: Table is adapted from \cite{abadie_samplingbased_2020}.}
    	\end{tabular}}
	\setbox0=\hbox{\contents}
    \setlength{\linewidth}{\wd0-2\tabcolsep-.25em}
    \contents}
\end{table}

In both approaches, uncertainty arises from missing data. In the sampling-based approach, the missing data are unobserved units in the population that did not make it into the sample. In the design-based approach, the missing data are the unobserved counterfactual for the observed unit. In making inference about an estimator, one could easily have a combination of both missing data problems – a subset of the population in which one observes only a single potential outcomes. As both approaches are essentially missing data problems, the estimated variance in the sampling-based approach is similar to the estimated variance in the design-based approach \citep{abadie_samplingbased_2020}. What differs is their interpretations.

If uncertainty comes solely from not observing all units in the population, the estimand is descriptive. If, however, uncertainty comes from unobserved potential outcomes, then the estimand is causal. As \cite{abadie_samplingbased_2020} write, ``articulating both the exact nature of the estimand of interest and the source of uncertainty that makes an estimator stochastic is a crucial first step to valid inference''. For this reason, we advocate full transparency in terms of writing explicitly what is being assumed about the model that leads one to select a particular inference approach. In writing papers, applied economists should formally state what is being assumed about the modeling or data generating process, why their chosen approach to inference is correct, and how and why that approach might fail (i.e., if the assumption is incorrect). Explicitly defining one’s assumptions and detailing how those assumption drive the analysis is common in economic theory but less common in applied economics. This sort of transparency would allow readers to place the empirical results in the context of those assumptions, and relative to other papers that make different assumptions. Ideally, these assumptions and choices would be pre-specified prior to the analysis, following the approach to writing and registering pre-analysis plans laid out in \cite{janzen_ulysses_2021}.

Once one has articulated the estimand and the source of uncertainty, the challenge is how to calculate the variance of the estimator when the sampling or assignment process is not from an \emph{i.i.d.} distribution. If this does not hold, then the distribution of $\epsilon$ will not satisfy the assumptions laid out above and calculating the variance of coefficients becomes more challenging. These challenges can arise when the errors are heteroskedastic, when errors are correlated within cluster or strata, when errors are serially correlated across time, or when one tests a family of hypothesis. We address each of these challenges to conventional inference using approaches that rely on traditional asymptotic theory, before turning to asymptotic refinements of each of these approaches.


\section{Challenges to Conventional Inference}

Traditional inference on cross-sections assumes data are \emph{i.i.d.} But, as any applied economist knowns, this is an unrealistic assumption. Even in a cross-section of independent draws, if the regression is a linear approximation of a nonlinear CEF, heteroskedasticity will exist \citep{angrist_mostly_2009}. In addition to heteroskedasticity, if data are grouped or stratified, errors may be clustered within the group or strata. Further, if there are panels of potentially unbalanced data, which provide multiple observations for the same unit, serial correlation can arise. Finally, if the hypothesis one is trying to make inference on is a member of a family of hypotheses, one may end up reject a single null hypothesis out of random chance. All of these cause bias in traditional inference.


\subsection{Heteroskedasticity}

A basic assumption of conventional inference is that the errors are homoskedastic, that is $\E[\epsilon_i^2 | X_i] = \Omega$, a constant. This assumption is imbedded in the default standard errors calculated by most software packages, including Stata. Yet, in most practical applications the assumption of homoskedasticity is unlikely to hold. Any time one fits a linear regression model on a nonlinear CEF, the residuals (the difference between the fitted line and the true nonlinear CEF) will vary with the size of values in $X_i$. Larger values in $X_i$ will yield larger residuals. Similarly, if the regression is a linear probability model (LPM), even if the CEF is linear, residuals will be heteroskedastic unless the regressor is constant \citep{angrist_mostly_2009}.

In large samples, we can rely on the Eicker-Huber-White (EHW) variance estimator, which is asymptotically consistent \citep{eicker_limit_1967, huber_behavior_1967, white_heteroskedasticity-consistent_1980}. This correction allows for the calculation of standard errors that are robust to heteroskedasticity of unspecified form. To get EHW robust standard errors, one simply estimates the OLS variance as:

\begin{equation}
    \V_{EHW} \left( \hat{\beta} \right) = \left( X_{i}^{'} X_{i} \right)^{-1} \left( \sum_{i} \Omega_{i} X_{i} X_{i}^{'} \right) \left( X_{i}^{'} X_{i} \right)^{-1},
\end{equation}

\noindent where $\Omega$ is no longer a constant but the variance of the estimated residuals:

\begin{equation}
    HC_{0}: \hat{\Omega}_{i} = \hat{\epsilon}_{i}^{2}.
\end{equation}

\noindent Here (and elsewhere in the literature) $HC$ stands for heteroskedasticity-consistent. Given the likelihood that the residuals from a regression model are heteroskedastic, one may be inclined to always calculate EHW robust standard errors. However, if the residuals are indeed homoskedastic, EHW standard errors will actually have a greater bias than conventional standard errors. So, one would not want to use robust errors unless one truly believes the residuals are heteroskedastic. Even when one uses the appropriate standard errors, both the conventional standard errors and the EHW standard errors can be asymptotically biased such that $\hat{\Omega}_i$ is too small \citep{angrist_mostly_2009}. Luckily, there are simple adjustments to correct for this bias. 

\cite{mackinnon_heteroskedasticity-consistent_1985} provide three possible corrections for bias in the EHW standard errors. Their first proposed correction is a simple adjustment for the degrees of freedom, similar to the finite-sample adjustment made to conventional standard errors in Stata and other software, discussed above. Instead of estimating $\hat{\Omega}_i = \hat{\epsilon}_{i}^{2}$, the formula is adjusted by the sample size $(N)$ and the degrees of freedom $(k)$:

\begin{equation}
    HC_1 : \hat{\Omega}_i = \frac{N}{N-k} \hat{\epsilon}_{i}^{2}
\end{equation}

\noindent For linear regressions in Stata, the $HC_1$ correction is the default method for calculating EHW standard errors. One defines the variance-covariance matrix as \texttt{vce(robust)}, and the degrees of freedom adjustment is built in.

$HC_1$ provides a conditionally unbiased estimate of the variance in the case that  residuals really are homoskedastic if all observations have the same leverage. Leverage is simply a measure of how far away an observation $x_i$ is from the mean. Formally, leverage is defined as:

\begin{equation}
    h_{ii} = h_{i}^{'} h_{i} = X_{i}^{'} \left( X^{'} X \right)^{-1} X_{i}.
\end{equation}

\noindent So, if all observations in the data are an equal distance from the mean, then $HC_1$ is the correct adjustment. However, if observations have different leverage, which they almost surely do, then the $HC_1$ adjustment does not completely correct for the bias.

The second proposed correction in \cite{mackinnon_heteroskedasticity-consistent_1985} is to adjust the variance based on the leverage of an observation. The adjustment for leverage is then:

\begin{equation}
    HC_2 : \hat{\Omega}_i = \frac{N}{1-h_{ii}} \hat{\epsilon}_{i}^{2}.
\end{equation}

\noindent By accounting for the leverage of each observation, $HC_2$ provides a conditionally unbiased estimate of the variance when the residuals really are homoskedastic. This adjustment can be implemented in Stata as \texttt{vce(hc2)}. While this adjustment is unbiased, \cite{angrist_mostly_2009} show that estimates still tend to be too small ``by accident'', though performance is better than $HC_1$ or conventional standard errors. The final proposed adjustment squares the denominator:

\begin{equation}
    HC_2 : \hat{\Omega}_i = \frac{N}{\left( 1-h_{ii} \right)^2} \hat{\epsilon}_{i}^{2}.
\end{equation}

\noindent While \cite{davidson_estimation_1993} report that this estimate performs better than the others when residuals really are heteroskedastic, the estimates will be biased if the residuals are homoskedastic. \cite{angrist_mostly_2009} show that the $HC_3$ correction tends to produce estimates that are larger than the true variance. This adjustment can be implemented in Stata as \texttt{vce(hc3)}.

In addition to the issue of bias in calculating EHW standard errors, there are two other issues. One issue is that the bias corrections proposed by \cite{mackinnon_heteroskedasticity-consistent_1985} increase the variability of the estimated variance. The other is the issue of what distribution to use in calculating test statistics. These issues are pronounced particularly in small samples. As \cite{angrist_mostly_2009} show through simulation, all four estimates of the variance (conventional and the three $HC$ corrections) result in rejection of the true null hypothesis too frequently. EHW standard errors can end up being smaller than conventional standard errors. In regard to this increased variability in the estimated variance, the practical result is that the tails of the empirical distribution are thicker than expected. This then raises the issue of what distribution to use in calculating test statistics. \cite{angrist_mostly_2009} show that rejection rates with EHW standard errors remain too high regardless of whether one calculates test statistics using a normal distribution or a $t$-distribution adjusted for degrees of freedom. They suggest a heuristic approach to this problem by estimating both the conventional and EHW standard errors and choosing the larger of the two. This ensures that applied economists take the conservative approach, accepting more null hypotheses than may be true (Type II errors).

Adjustments to the EHW standard errors, and the issues associated with these adjustments, is typically thought of as a problem only in small samples. As $N \rightarrow \infty$, the large sample properties mean that the adjustments for degrees of freedom or leverage have a diminishing impact on the calculation of the variance. To understand why, one needs to recall that robust covariance matrices are conditionally biased, because the residuals are downward biased estimates of the regression errors. The latter are what go into the (asymptotic) covariance matrix of the OLS parameters while the former just go into the estimator. The bias gets larger as leverage gets larger. Since leverage must sum to the dimension of the covariate matrix, everyone’s leverage goes to zero as $N \rightarrow \infty$, meaning the bias diminishes as the sample size gets larger.

However, as \cite{imbens_robust_2016} point out: problems with the EHW standard errors can be pronounced in samples of $50$ observations or $50$ clusters. So, if one is analysing the impact of differences in school lunch policies by U.S. states, bias in EHW standard errors are of concern. This is because the problem with EHW standard errors is not just the bias, but excessively large variability. The accuracy of EHW standard errors is a combination of sample size and leverage. Instead of using one of the \cite{mackinnon_heteroskedasticity-consistent_1985} corrections, or the \cite{angrist_mostly_2009} heuristic, \cite{imbens_robust_2016} recommend an adjustment first proposed by \cite{bell_bias_2002}, hereafter BM. The BM correction adjusts not only for the bias in the EHW standard errors, but the distribution used for calculating test statistics. The bias adjustment in BM is simply the $HC_2$ correction. Then, instead of using a normal distribution or the degrees of freedom adjusted $t$-distribution, BM uses an adjustment that matches the moments of the variance estimator to one of a $\chi^2$-distribution. This approach has been refined by \cite{tipton_small-sample_2015} and \cite{pustejovsky_small-sample_2018}. The BM correction to the EHW standard errors can be implemented in Stata using the user-written command \texttt{reg\_sandwich} \citep{tyszler_reg_sandwich_2017}. \cite{imbens_robust_2016} recommend that applied economists routinely estimate EHW standard errors using the BM correction.

Most work on developing and testing refinements to EHW standard errors use simulated data and the focus is frequently on the Behrens-Fisher problem. This is the case in both \cite{angrist_mostly_2009} and \cite{imbens_robust_2016}. The Behrens-Fisher problem, as formulated by \cite{deaton_instruments_2010}, is the problem of testing for the difference between the means of two populations (a treated population and a control population) that have different variances. The problem is typically formulated using a special case of linear regression with a single binary regressor:

\begin{equation}
    y_i = \alpha + \beta t_i + \epsilon_i .
\end{equation}

\noindent When the variances of the two populations differ, meaning errors are heteroskedastic, the exact finite-sample distribution is unknown. The \cite{mackinnon_heteroskedasticity-consistent_1985} corrections and the BM correction provide asymptotic approximations to this unknown distribution.

The use of the Behrens-Fisher problem to produce corrections for heteroskedastic errors has created its own problem. \cite{young_improved_2016} finds that for a third of the $1,378$ regressions used in the $44$ published papers he investigates, the \cite{mackinnon_heteroskedasticity-consistent_1985} and BM refinements are infeasible. This is because the regression design used in these papers is such that the regressions having a maximal leverage of one, meaning the commonly promoted correction requires inverting a singular matrix. While the refinements work for Behrens-Fisher-like models, they are unusable in a great many applied economics papers. \cite{young_improved_2016} examines regressions from 44 experimental papers published in the journals of the American Economic Association. He points out that a quarter of the regressions in these papers have a leverage of one, meaning again that the BM and other refinements are infeasible. Because the limitation to these refinements is poor regression design, \cite{young_improved_2016} proposes an effective degrees of freedom (EDOF) correction based on what actually is included in the regression to eliminate bias and adjust the $t$-distribution so that the variance of the test statistic does not inflate. This adjustment is similar to the BM adjustment, though those details are beyond the scope of this paper. \cite{young_improved_2016} demonstrates that the bias and EDOF correction yields nearly exact inference under homoskedasticity and improvements under heteroskedasticity in terms of both standard errors and rejection rates. The adjustment is effective for all three corrections proposed by \cite{mackinnon_heteroskedasticity-consistent_1985}.

The most helpful aspect of the Young adjustment is that is reduces the bias in EHW standard errors even when residuals are indeed homoskedastic. Thus, one can always calculate Young-adjusted EHW standard errors, knowing that if the errors are homoscedastic, Young's adjustment will yield nearly exact standard errors in this case. The takeaway is that one can always calculate robust standard errors, with the EDOF correction simply by using the user-written \texttt{edfreg} command in Stata \citep{young_edfreg_2016}. The Young EDOF correction should be given preference to the $HC_1$ correction of the EHW standard errors that is common in the literature and the default in Stata, as well as many other statistical packages.


\subsection{Clustering and Serial Correlation}

Beyond the assumption of homoskedasticity, it is traditional in applied economics research to assume that the data are independent, meaning that each observation is a random draw from some population. This independence assumption means that the residuals in a regression will be uncorrelated with each other. However, most microdata is clustered, grouped, or stratified, either from sampling method or from the research design of an experiment \citep{abadie_samplingbased_2020}. Clusters may be villages in which an experimental treatment was delivered, or they may be all the states, counties, or provinces within a country, or the sampling process may be stratified to ensure representation of certain groups. In any of these cases, if the assignment to treatment or the sampling process varies by group, or if clusters are sampled randomly from a larger population of clusters, the independence assumption will not hold. As with clustered data, longitudinal data have become much more common in applied microeconomics. Even in completely randomized experiments, it is now standard practice to collect pre- and post-treatment data to allow for difference-in-difference or ANCOVA estimation. Repeated sampling from the same unit over time gives rise to serial correlation, meaning residuals within the unit will be correlated.

In the program evaluation literature, the issue of clustering was brought to prominence by \cite{bertrand_how_2004}. The authors survey papers that use difference-in-difference and find that less than half (36 out of 80) of the papers address the problem of correlated errors in their grouped data. Similarly, only five of 69 papers with more than two periods of data address the issue of serial correlation. While clustering is now much more common, misconceptions persist on when clustering is necessary \citep{abadie_when_2017}. \cite{cameron_practitioners_2015} focus on clustering as a data issue, arguing that if individual observations are correlated within groups, then the practitioner needs to correct the standard errors for this cluster-based correlation. However, this empirical approach means that it can be difficult to know at which of the potentially many group levels one should cluster. Should one correct for correlation at the individual-level, the village-level, the state-level, or the state-year level? Residuals are likely to be correlated within all these groups. As an alternative, \cite{abadie_when_2017} argue that clustering is a design issue and should be justified by the design (either sampling or research) of the study. They further argue that other justifications (correlation in regressors, correlation in residuals, that the cluster correction has an effect) are likely to lead applied economists astray.

The standard correction for within-group correlation was developed by \cite{liang_longitudinal_1986}, hereafter LZ, who generalize White’s (1980) robust covariance matrix to allow for grouped data. This correction allows for the calculation of standard errors that are robust to within-cluster correlation of unspecified form. To get LZ cluster-robust standard errors, one simply estimates the OLS variance as:

\begin{equation}
    \V_{LZ} \left( \hat{\beta} \right) = \left( X_{i}^{'} X_{i} \right)^{-1} \left( \sum_{c} X_{c}^{'} \Omega_{c} X_{c} \right) \left( X_{i}^{'} X_{i} \right)^{-1}
\end{equation}

\noindent where $\Omega_c$ is a submatrix of corresponding units from cluster $c$. To get consistent estimates of the variance, we use the estimated residuals:

\begin{equation}
    \hat{\Omega}_{c} = a \hat{\epsilon}_{c} \hat{\epsilon}^{'}_{c}
\end{equation}

\noindent where $\hat{\epsilon}_c$ is a column vector and $a$ is a degrees of freedom adjustment similar to that in $HC_1$. This is the calculation done by the Stata command using \texttt{vce(cluster clustvar)}. It corrects for within-cluster correlation of residuals without placing any restrictions on the nature of the correlation. This means that the LZ standard errors will correctly adjust for correlation regardless of whether it comes from grouped data or serial correlation.

Monte Carlo simulations in \cite{angrist_mostly_2009} show that adjustments for heteroskedasticity tend to have only a small effect on inference, no more than 30 percent in the size of the standard errors. However, clustering can lead to large changes in standard errors, resulting in large differences in $t$-statistics and $p$-values. As with the EHW standard errors, in recent years there have been several refinements proposed to the LZ standard errors. \cite{imbens_robust_2016} recommend the \cite{bell_bias_2002} version of the cluster correction to the LZ formula. As discussed above, the BM correction is both a bias and degrees of freedom adjustment, generalized to grouped data, similar to $HC_2$. But the same problems with using the BM correction for bias in EHW standard errors hold true for the BM correction to LZ standard errors. As \cite{young_improved_2016} points out, the BM correction is infeasible in $38\%$ of regressions in his database of 44 experimental studies published in the AEA journals. Young’s (2016) EDOF correction for EHW standard errors also works to correct LZ standard errors and can be implemented in Stata with his \texttt{edfreg} command.

Cluster correction is straightforward in large-sample, cross-sectional data. However, the approach breaks down if we have few groups or if we have grouped data over time. The issue of ``too few'' groups is analogous to the small sample issues discussed in relation to heteroskedasticity. Fundamentally, the LZ correction relies on asymptotic theory, requiring for consistency that the number of clusters tends towards infinity. When the number of clusters is fixed, such as the number of member states in the European Union, then $\V_{LZ} \left( \hat{\beta} \right)$ is not consistent, even if the number of observations within each cluster tends towards infinity. This is because the sums are taken across $c$ (clusters), not across observations, $n$ \citep{angrist_mostly_2009}. A further issue exists if the grouped data are observed over time since it requires the researcher to know how to deal with serial correlation at the individual-level when there are grou$p$-time shocks. A classic example is in difference-in-difference models where treatment is randomized at the level of the group \citep{bertrand_how_2004, cameron_practitioners_2015}. In this case, there is within-cluster correlation of the regressors (group assignment to treatment) and also within-cluster correlation of the residuals (serial correlation). An example would be the impact of different state-level industry regulations on firm-level production costs. There will be both within-state correlation of regressors and correlation of firm-level residuals. The questions then become: how many clusters is ``too few'', and, at what ``level'' should one adjust for clustering?

In answering the question about ``too few'' clusters, Angrist and Pischke again provide a helpful heuristic. They suggest that 42 clusters are sufficient for reliable inference. In a similar vein, \cite{ozler_beware_2012} writes ``in practice, having 30 to 40 clusters is like approaching infinity''. However, \cite{abadie_when_2017} take issue with the reliance on empirically driven justifications for using asymptotically derived inference, if, in fact, the number of clusters is fixed. \cite{cameron_practitioners_2015} highlight the point that there is no theory or rule regarding what constitutes enough clusters. If the number of observations in each cluster is equal across clusters (the clusters are balanced), then one may need as few as 20. But, if the number of observations varies across cluster, the effective number of clusters decreases \citep{imbens_robust_2016}. Further complicating the issue, \cite{young_improved_2016} demonstrates that the leverage also influences the effective number of clusters. In other words, it is not just balance in cluster sizes that matters, but balance in the covariance matrix. Thus, a researcher doing inference on data with $50$ balanced clusters may be acceptable, while a researcher with unbalanced clusters will need more clusters \citep{carter_asymptotic_2017}. The Stata command \texttt{clusteff}, developed by \cite{lee_multiple_2014}, checks for the severity of heterogeneity in cluster size and leverage and then calculates the effective number of clusters.

In terms of at what ``level'' one should cluster, several options are available. \cite{bertrand_how_2004} and \cite{angrist_mostly_2009} compare a set of alternative methods to deal with grouped panel data. One option, which tends to perform poorly, is to use parametric methods to try and model the serial correlation \citep{bertrand_how_2004}. A second, and better, solution is to collapse, by averaging, the data into two time periods. In a difference-in-difference context, this involves taking time averages of outcomes prior to the treatment and then time averages of outcomes after the treatment. One can then consistently estimate standard errors without concern of serial correlation in the residuals. While this approach performs well, it comes at the cost of a loss of power, as it effectively reduces the time dimension, $t$, to two \citep{mckenzie_beyond_2012}. A third option is to cluster at the intersection of the two dimensions: group $\times$ time. However, this approach makes the implicit assumption that the residuals in group $A$ at time $t$ are uncorrelated with the residuals in group $A$ at time $t+1$ (or any other time). Cameron and Miller (2015) propose a better solution, which is to implement multiway clustering, effectively accounting for both within-cluster correlation in the regressions (grouped data) and within-cluster correlation in the residuals (serial correlation). Multiway clustering can be implemented using \texttt{emgreg} or \texttt{ivreg2}, both user-written commands \citep{baum_ivreg2_2002, cameron_robust_2011, baum_enhanced_2007}. The commands calculate the cluster-robust variance-covariance matrix at the first level (group), then at the second level (time), then at the intersection of the two (group $\times$ time). They then add the first two and subtract off the third. While this approach allows for clustering along two dimensions, it requires a ``large enough'' number of clusters in both levels.

In general, the approach to the key issues of clustering (i.e., the number and the level) has been sample-based, relying on a set of heuristics and evidence from the data itself. \cite{abadie_when_2017} suggest that the common view among applied economists seems to be that if LZ standard errors are different than normal standard errors, then one should cluster. They argue that a design-based approach can provide clarity on issues of when to cluster, particularly at what level clustering should occur. This approach calls for the researcher to think carefully, and define clearly for the reader, how the sample was collected and/or how the experiment was designed. Specifically, for randomized experiments, one should cluster at the level of treatment assignment. If treatment and outcome are measured at different units, such as village-level treatment and farmer-level outcomes, and if there is a time dimension to the experiment, then one should use the multiway clustering proposed in \cite{cameron_practitioners_2015}. In non-experimental situations, clustering at the unit-level will typically be necessary in panel data to correct for serial correlation. As an example, in a state level difference-in-difference, state-level clustering will be necessary if $t$ is serially correlated, as it will be if a policy ``turns on'' and then stays on. Applied economists should provide justification for why they chose to cluster and at what level, based on the design of the sample and the experiment. A design-based approach will provide greater direction for the researcher and greater insight for the reader than the \emph{ad hoc} approach common today.


\subsection{Multiple Inference}

The challenges to inference from heteroskedasticity, grouped data, and serial correlation are well known and documented in the economics literature. Less is said, however, about the role that multiple hypothesis testing plays in inference. When applied economists seek to test a number of hypotheses with respect to a single theory, coefficients can emerge as significant simply by chance – resulting in a Type I error. As an example, suppose one wanted to test if a policy had a statistically significant effect at reducing food loss and waste, measured in two different ways. Since one is performing two independent tests, the probability of not making a Type 1 error when $\alpha = 0.05$ is $0.95 \times 0.95 = 0.9025$. The probability of concluding there is a statistically significant effect, when there is in fact no effect, is $1 - 0.9025 = 0.0975$ instead of $0.05$. By testing two hypotheses, the researcher is giving themselves multiple chances to reach a conclusion. This fact needs to be accounted for when making inference.

Depending on what one is interested in proving, testing multiple hypotheses can give rise to either the family-wise error rate (FWER) or the false discovery rate (FDR). The FWER is the probability of making at least one false discovery among a family of comparisons, as in the example above. The FDR is the probability of making at least one false discovery among the discoveries already made. The FDR accounts for dependencies among hypotheses by correcting for the expected proportion of rejections that are Type I errors. Thus, the FDR is frequently less conservative than the FWER and is preferred when the number of statistical tests being performed is substantial. None of these key terms or concepts appears in classic microeconometric textbooks like \cite{cameron_microeconometrics_2005} or \cite{wooldridge_econometric_2010}. Even in recent textbooks focused on causal inference, such as \cite{imbens_causal_2015} and \cite{morgan_counterfactuals_2015}, the problem of multiple hypothesis testing is not addressed.

Although the problem of multiple comparisons is relatively new to applied economists, the problem itself is well known in other fields. In fact, the problems arising from multiple inference dates back to \cite{bonferroni_il_1935}, who first proposed a simple correction for the FWER. Suppose one wanted to test the effectiveness of two different treatments on five different outcomes. The result is ten hypotheses being tested. If none of the treatments have any effect, meaning the nulls are true, using a critical value of $\alpha = 0.05$, there would be a $1 - 0.95^{10} = 40\%$ probability of one or more false rejections. The probability of Type I errors rises with larger critical values $\left( \alpha = 0.10 \right)$, with more treatments, and with more outcomes. \cite{bonferroni_il_1935} suggests reducing the critical value based on the number of hypotheses being tested, making it more difficult to reject the null. In our example, the Bonferroni correction would require each hypothesis to be tested against the critical value $\alpha = 0.05 \slash 10 = 0.005$ for it to be rejected. While simple, the Bonferroni correction can be overly conservative when there are a large number of hypotheses or hypotheses are correlated \citep{list_multiple_2019}.

The economics literature has yet to arrive at a consensus on the best way to correct for multiple hypothesis testing \citep{viviano_when_2021}. Some suggest adjusting only when making inferences for multiple outcomes \citep{anderson_multiple_2008, casey_reshaping_2012, heckman_inference_2011, kling_experimental_2007} while others suggest correcting only for multiple subgroups \citep{lee_multiple_2014}. Still others suggest correcting for both multiple outcomes and subgroups \citep{heckman_analyzing_2010}. There have also been numerous recent proposals for more efficient, more powerful, and less conservative methods to correct for either the FWER or the FDR \citep{romano_hypothesis_2010}. Many of the approaches rely on a step-down procedure first proposed by \cite{holm_simple_1979}. Assume one is testing a family of $m$ hypotheses and wants to ensure the FWER will be no higher than $\alpha = 0.05$. Start by sorting the $p$-values for each hypothesis from smallest to greatest. Then, compare the first $p$-value to the Bonferroni corrected critical value. If $p_1 < \alpha \slash m$, then reject the corresponding hypothesis and proceed to the next largest $p$-value, comparing $p_2 < \alpha \slash (m-1)$ and so on until one reaches the first non-rejected null. All hypotheses with $p$-values larger than the $p$-value on the first non-rejected hypotheses are also not rejected. The Holm correction is uniformly more powerful than the Bonferroni correction, meaning that it is able to correctly reject false null hypotheses at least as well as the Bonferroni correction. That said, the Holm correction is still conservative, meaning it is more likely to commit a Type II error, when hypotheses are not independent.

\cite{westfall_resampling_1993} propose a step-down method with more power, because it allows for dependency between $p$-values. The Westfall-Young correction uses a permutation procedure (bootstrap) to estimate the dependence relationship between hypotheses. This dependence structure is then used to adjust the step-down process, providing uniformly more power than the Holm correction, meaning adjusted $p$-values are at least as small with Westfall-Young as with Holm. The procedure assumes subset pivotality in order reduce the number of comparisons between hypotheses necessary and thereby simplify the computational process \citep{westfall_multiple_2008}. The Westfall-Young procedure is asymptotically optimal \citep{meinshausen_asymptotic_2011} and can be implemented with the user-written Stata command \texttt{wyoung} \citep{reif_wyoung_2017}.

\cite{romano_exact_2005, romano_stepwise_2005} propose a generalization of the Westfall-Young correction. Like \cite{westfall_resampling_1993}, the Romano-Wolf correction uses a bootstrap resampling procedure to allow for dependence across outcomes. The procedure uses the dependence among $p$-values to increase power, meaning it has greater ability to detect genuinely false null hypotheses. Unlike the Westfall-Young procedure, the Romano-Wolf correction does not use the subset pivotality assumption to simplify computation \citep{clarke_romano-wolf_2019}. The procedure can be implemented in Stata via the user-written command \texttt{rwolf} \citep{clarke_rwolf_2016}. Building on \cite{romano_balanced_2010}, \cite{list_multiple_2019} develop a FWER correction that relies on a similar bootstrap resampling procedure. \cite{list_multiple_2019} make an additional assumption that assignment to treatment is at the individual level and done through simple random sampling. For applied economists whose data satisfies this assumption, the List-Shaikh-Xu correction provides more power than either the Bonferroni or Holm correction. However, it is not clear what improvement, if any, the List-Shaikh-Xu correction makes on Westfall-Young or Romano-Wolf. \cite{seidel_mhtexp_2016} provide code to implement the List-Shaikh-Xu correction in Stata via the command \texttt{mhtexp}. However, the command itself is rather limited in that it does not allow for different controls in different regressions. \cite{steinmayr_mhtreg_2020} provides a more generalized version of the \texttt{mhtexp} command, called \texttt{mhtreg}, which allows for the inclusion of different controls in different regressions, as well as for clustered randomization.

Most of the recent developments in multiple inference have been to corrections for the FWER and not the FDR. This may be due to the fact that FDR techniques are less conservative, meaning they are more likely to reject a true null due to random chance from testing multiple hypotheses. In general, applied economists prefer to be conservative when rejecting null hypotheses, preferring to commit Type II errors to Type I errors. However, the FWER corrections become increasingly severe as the number of hypotheses being tested grows. As experiments become ever more complex, with multiple treatment arms and multiple outcomes, the FWER corrections may be too conservative. This is because the procedures to control for the FWER limit the probability of making \emph{any} Type I error. Alternatively, controls for the FDR seek to limit the proportion of Type I errors. While the approach is less conservative, it generally provides more power than the FWER approach to controlling for multiple inference.

\cite{benjamini_controlling_1995} propose a simple step-up procedure to control the FDR. Assume one wants to ensure the FDR will be no higher than $\alpha = 0.05$. As with \cite{holm_simple_1979}, order the $p$-values from smallest to greatest. Assign each $p$-value a rank $i$, so that the smallest $p$-value is $i = 1$, the second smallest is $i = 2$. Then calculate the Benjamini-Hochberg correction for each $p$-value as $q = (i \slash m) \alpha$. These corrected $p$-values are known as $q$-values. Find the largest $p$-value such that $p \leq q$ and reject the corresponding null hypothesis and all other null hypotheses which ranked higher (closer to $i = 1$) than the first rejected hypothesis. The step-up process ensures that the FDR is controlled for at whatever $\alpha$ is chosen and is slightly conservative in rejecting nulls.

\cite{benjamini_adaptive_2006} provide a two-stage procedure to calculate sharpened $q$-values, which are more accurate than just calculating $q$-values. The first stage is just the Benjamini-Hochberg procedure. In the second stage, $q$-values are recalculated using the number of hypotheses rejected in the first stage. These sharpened $q$-values provide better power when $p$-values are independent or positively dependent (Benjamini et al., 2006). While there is not a package to calculated sharpened $q$-values, \cite{anderson_multiple_2008} provides code for the calculations, which is easily implemented in Stata. Sharpened $q$-values can be smaller than uncorrected $p$-values if there are many true hypotheses, because in these cases controlling for the FDR can allow for some false rejection while still maintaining a given $\alpha$. This is a key difference from FWER corrections, which seek to limit \emph{any} Type I error.

In general, one should correct for multiple inference any time one is testing multiple hypotheses related to one theory. In applied economics, correction for multiple hypothesis testing has typically been limited to experimental studies, either in the lab or in the field. But the problem of multiple inference is not a function of whether the data are experimental as opposed to observational. It is unclear why expectations in the profession for adjusting for multiple inference differ based on the data generating process. If a researcher randomized the provision of hybrid maize seed to smallholder farmers in Kenya and then collected multiple measurements of welfare outcomes, it would be standard practice to adjust critical values to account for the greater probability of making a Type I error. Yet, if adoption of the hybrid maize seed were through self-selection instead of randomization, the researcher would not be expected to correct for multiple inference. We believe that any time applied economists give themselves multiple chances to reach a conclusion by testing the impact of a variable on an outcome measured in multiple ways, they should correct for the Type I error probability. When there are a limited number of hypotheses one should control for the FWER and when there are many hypotheses the less conservative FDR can be used. What qualifies as a few or many hypotheses remains an open question and should be justified as part of the research design or pre-analysis plan.


\section{Asymptotic Refinements to Conventional Inference}

As the cost of computation has declined, asymptotic refinements based on numerical methods have become more common \citep{mackinnon_bootstrap_2006}. Chief among them is the bootstrap, a term coined by \cite{efron_bootstrap_1979, efron_jackknife_1982}. The term comes from the idiom, ``to pull yourself up by your own bootstraps'', meaning to use one’s existing resource to improve one's situation. By re-sampling from existing data in such a way as to mimic that underlying data, the bootstrap can provide refinement on population statistics. Bootstrapping offers an alternative to conventional inference, which relies on asymptotic formulas, and in many cases, bootstrapping improves on asymptotic approximations to the unknown exact finite sample distribution \citep{horowitz_bootstrap_2019}. A bootstrap sample is drawn from one’s own data, in which the original sample is treated as if it were the population. One repeatedly draws a new random sample (with replacement), which are the bootstrap samples. The bootstrap distribution is the distribution of these many samples. We expect that the bootstrap distribution constructed by drawing from one’s own data will provide a good approximation of the sampling distribution.

The value of bootstrap sampling methods is that under mild regularity conditions they can yield approximations to test statistics as good as or better than those provided by asymptotic theory \citep{davidson_bootstrap_2006}. Further, they require fewer assumptions and restrictions than alternative methods, such as the delta method. This has always been the great promise of the bootstrap: greater precision with fewer assumptions. As the type and variety of data used by applied economists increases (e.g. panel data, randomized experimental data, data on an entire population) the assumptions necessary for invocation of asymptotic theory become less reasonable. Additionally, the sample size may be too small, or the number of clusters in the data too few, for believable reliance on asymptotic theory. In many cases, applied economists can no longer rely on the sampling-based uncertainty which underpins conventional asymptotic inference. Until recently, the bootstrap’s promise had been constrained by the available computing power and the complexity of writing one’s own bootstrap program. However, in the last decade, these constraints have been relaxed and most statistical software makes bootstrapping relatively simple to implement. But, understanding when to use these methods, and which of a number of alternative methods to select, remains beyond many practitioners.

We advocate for three things. First, that most applied economics research should rely on asymptotic refinements to conventional test statistics when conducting inference. As \cite{horowitz_bootstrap_2019} writes, ``the ability of the bootstrap to provide asymptotic refinements for smooth, asymptotically pivotal statistics provides a powerful argument for using these statistics in applications of the bootstrap whenever possible''. That is to say, in almost all cases, the main results in a paper should rely on inference based on appropriately calculated bootstrapped test statistics and not those that rely on first-order asymptotic theory. Obviously, there remain instances where asymptotically derived test statistics will be superior to bootstrapped statistics. However, we believe these instances are a minority of cases. Applied economists should justify why they do not need to bootstrap test statistics as opposed to the reverse. As such, inference that relies on asymptotic approximations can still be presented, but they should be presented as ``robustness checks'', in the way that bootstrapped results are typically presented now. Second, applied economists need to be using many more bootstrap replications $(r)$ than is currently common. Researchers frequently use somewhere in the range of $r = 500$ to $r = 1,000$ bootstrap replications. \cite{hesterberg_what_2015} shows that in many applications, researchers should be using $r \geq 5,000$ replications for bootstrapping their standard errors and $r \geq 10,000$ or $r \geq 15,000$ for pivotal statistics, such as $t$-statistics, depending on how much accuracy matters. Third, while extremely powerful, bootstrapping is not a panacea for all challenges to conventional inference. A small sample, for example a sample with $n = 50$,  is unlikely to represent a large population well. Bootstrapping will not change this fact, meaning it cannot overcome the weaknesses of small samples as a basis for inference \citep{hesterberg_what_2015}.

The primary hurdle implementing these recommendations has historically been a lack of bootstrap packages and the time-consuming nature of creating one’s own bootstrapping program. While this has been largely addressed through statistical software and user-written packages, there still does not exist a bootstrap package for all applications: that is, there is no package that can compute asymptotic refinements while accommodating all of the challenges to classical inference discussed throughout this chapter. In what follows, we outline three bootstrap methods and how they can help in addressing the challenges to conventional inference already discussed.


\subsection{Classic Bootstrapping}

The classic bootstrap method is generally referred to as a ``pairs'' or ``nonparametric'' bootstrap. Classic bootstrap methods are extremely flexible and typically require only mild regularity conditions on the CEF. Assuming the data are independent draws from a distribution, and the CEF is a smooth function, classic bootstrapping can improve on asymptotic approximations to pivotal statistics.

In the classic bootstrap, pairs of the dependent variable and the covariate values are drawn to bootstrap the regression estimates. This method is most appropriate in situations where one does not wish to guess at the distributional form, as the classic bootstrap makes no assumptions about the functional form distribution. Starting with an original sample of $n$ observations, one draws at random a pair $(y_i,X_i )$, the dependent variable and its corresponding covariates. This pair is then saved as the first observation in a new bootstrap sample. The pair is then replaced in the original sample and a new pair is drawn at random and placed in the bootstrap sample. The process is repeated $n$ times. Because the bootstrap sample randomly draws with replacement from the original sample, some original observations will appear multiple times while others will not appear at all. One then calculates the mean and variance of the coefficients of interest, and any test statistics, and saves them. The process is then repeated again and again to build the bootstrap distribution of replicated statistics.

In the case of bootstrapping standard errors, these are calculated as

\begin{equation}
    \hat{\sigma} = \bigg\{ \frac{1}{r-1} \sum \left( \hat{\beta} - \bar{\beta} \right) ^{2} \bigg\}^{1/2}.
\end{equation}

\noindent Here $r$ is the number of replications, $\hat{\beta}_i$ is the estimated coefficient from the $i^{th}$  bootstrap sample, and $\bar{\beta}$ is the mean value of the coefficient in the bootstrap distribution. While the mean of the bootstrapped coefficient is used in calculating standard errors, the coefficient calculated from the original sample is retained as the unbiased point estimate and the bootstrap is used just to calculate standard errors or other test statistics for inference.

For bootstrapped confidence intervals, the procedure is known as percentile bootstrapping. Having built the bootstrap distribution, one then orders $\hat{\beta}{_i}$ such that $\hat{\beta}_{1} \leq \cdot \cdot \cdot \leq \hat{\beta}_{r}$. For constructing a two-sided $95\%$ confidence interval, the lower and upper confidence bounds are the ordered elements, respectively. So, for $r = 10,000$, these would be the $250^{th}$ and $9,750^{th}$ ordered elements.

To generate bootstrapped $t$-statistics, one calculates:

\begin{equation}
    t_{i} = \frac{\hat{\beta}_{i} - \bar{\beta}}{\hat{\sigma}_{i}}
\end{equation}

\noindent where $t_i$ is the $t$-statistics for the $i^{th}$ bootstrap sample. Order $t_i$ by the absolute value, such that $| t_1 | \leq \cdot \cdot \cdot \leq | t_r |$, which ensures symmetric critical values. One then selects the appropriate critical value from these ordered elements. So, for a two-sided $95\%$ test, and $r = 10,000$, the ordered element is the $9,500^{th}$. One can also use this procedure to create bootstrapped confidence intervals for hypothesis testing instead of the ones from the standard normal tables.

In the case of moderate heteroskedasticity, classic bootstraps are appropriate to provide an inference strategy with bias-corrected standard errors and a minimal loss of precision in estimates \citep{angrist_mostly_2009}. Using classic bootstraps, inference based on asymptotic theory can be improved upon because the bootstrap provides bias reduction in finite sample estimators \citep{horowitz_bootstrap_2001}. Additionally, the classic bootstrap can provide asymptotic refinement to asymptotically pivotal statistics, such as the $t$-statistic, as well as improvements on hypothesis tests and confidence intervals based on the $t$-statistic. This is because the error in the bootstrap’s approximation of the distribution of the test statistic converges to zero faster than the error in the asymptotic approximation.

Classic bootstrap methods are relatively straightforward for estimation using Stata. The simplest way to generate bootstrapped standard errors is to use \texttt{vce(bootstrap)}. This is recommended over other programs for any estimation which allows the command because it automatically handles clustering and other model specifics. However, other commands are available, including \texttt{bsample}, which draws a sample with replacement from a dataset and the bootstrap prefix. With the \texttt{bootstrap} prefix, users can use bootstrap methods to estimate a function of stored results of existing commands. This allows for estimation of bootstrapped confidence intervals and $t$-statistics. We believe this method should be the standard in applied economics, following \cite{horowitz_bootstrap_2019}, who recommends bootstrapping test statistics for hypothesis testing and constructing confidence intervals, not simply bootstrapping standard errors.


\subsection{Residual and Wild Bootstrapping}

Although classic bootstrap methods are the most common, they are not appropriate in all circumstances. The classic bootstrap makes no assumptions about the functional form of the distribution, though it does make a somewhat strong assumption that the bootstrap samples are from the same joint distribution. Alternative methods make different assumptions regarding distributions and functional forms. Residual-type bootstrapping is one of these alternatives.

In residual bootstrapping, covariate values are fixed, and one draws samples from the distribution of residuals to generate a new estimate of the dependent variable, based on the predicted values and residual draws for each observation. As with classic bootstrapping, this process is repeated to build a residual bootstrap distribution. This method mimics a sample drawn with non-stochastic errors and ensures that the covariates and regression residuals are identically distributed. However, this latter assumption is not helpful if one is interested in standard errors under heteroskedasticity, which, as discussed, is likely to be the case with data of interest to applied economists.

Wild bootstrapping is a method which relaxes the independence assumption to allow for heteroskedasticity. \cite{wu_jacknife_1986} and \cite{liu_bootstrap_1988} first proposed wild bootstrapping which \cite{cameron_bootstrap-based_2008} adapted to for use when residuals are serially correlated or when there is within-cluster correlation. Wild bootstraps effectively match the unknown exact finite sample distribution up to a limited number of moments, making inference more reliable \citep{mackinnon_bootstrap_2006}. Wild bootstrap methods presume that the ``true'' residual distribution is symmetric and can provide advantages over residual bootstrapping, in particular for smaller sample sizes or small number of clusters. 

In the wild bootstrap, as in the residual bootstrap, covariates are held fixed and one resamples the dependent variables based on residual values. For each draw, one computes a new dependent variable, such that the residuals are multiplied by a random variable, with mean of zero and variance of one. This random variable is sometimes called a ``wild weight'' \citep{roodman_fast_2019}. As with other bootstrapping methods, this process is repeated to build a bootstrap distribution which is used for inference. Wild bootstrapping is straightforward using the user-written Stata command \texttt{boottest} \citep{roodman_boottest_2015}. This command uses a percentile $t$-bootstrap. In this process, one first subtracts the mean from each observation in the sample, so that the mean of the centered distribution is zero. Next, one samples with replacement from this centered distribution and, for each random sample, a $t$-statistic is calculated as in the classic bootstrap. From this, one obtains the bootstrapped distribution of the $t$-statistic expected by random sampling, under the hypothesis that the population has a mean of zero, given the distribution of the data. \cite{roodman_boottest_2015} discuss the procedure in more detail, including the various regression commands for which \texttt{boottest} is appropriate as a post-estimation method.

The takeaway regarding both classic, residual, and wild bootstrapping is that they provide a straightforward method to estimate standard errors, confidence intervals, and $t$-statistics, providing more accurate estimates than standard statistics obtained using asymptotic approximations. We advocate that applied economists should primarily rely on asymptotic refinements to conventional test statistics when conducting inference and that the main results in a paper should rely on inference based on appropriately calculated bootstrapped values. While bootstrapping cannot solve every problem in making inference, the procedures are sufficiently general enough that they can and should be applied in most applications.  


\subsection{Randomization Inference}
A distinct though related numerical method for inference is randomization inference (RI), alternatively called permutation tests. RI provides a method for calculating the significance of estimates, accounting for variations that may arise from randomization itself. That is, RI considers not just what would have happened under random assignment to some treatment but under all possible assignments. This generates a null distribution which can be compared to the actual estimate. RI remains infrequently used in applied economics, though with recent programming innovations, the costs for undertaking such analysis have rapidly declined \citep{hes_randomization_2017, young_asymptotically_2020}. The RI method was pioneered by \cite{fisher_design_1935} and extended by \cite{rosenbaum_observational_2002}. The method has typically been applied to the analysis of randomized control trials (RCTs), where randomization is built into the design of the study, or to studies where the entire population is observed \citep{ganong_permutation_2018}. Yet a designed-based approach to inference means that RI methods have widespread potential applications in applied economics.

In applied economics, most inference is built on sampling-based uncertainty, assuming a large sample drawn from an infinite population, and relies on the asymptotic traits of the estimator. However, this assumption becomes less appropriate as the types of data available to applied economists expands. RI is a useful technique when the method of randomization is known, when there are a small number of observations, when the entire population is observed, or when there is a high degree of leverage \citep{young_channeling_2019}. In any of these cases, RI can determine whether the observed outcomes in the data are likely to have been observed by chance, even if the treatment had no effect.

The steps for RI are conceptually straightforward in a Monte Carlo framework. To fix ideas, assume one runs a simple, completely randomized experiment by randomizing 5,000 consumers into either a treatment, where they see calorie counts on a restaurant menu, or a control group, where calorie counts are not provided, with probability $0.5$ on each. At the end of the experiment, one will have an observed outcome $y_i$, the calories on the ordered menu item, for each person in the study. One can then estimate the effect of the treatment, $x_i$, on outcome via a simple linear regression. To conduct randomization inference, one needs to construct a RI $p$-value. To do this, first save the original treatment status. Second, re-do the randomization process to get a new, randomly assigned treatment for each consumer. This new fake treatment status is attached to the consumer’s real outcome. Run the exact same original regression as before but now with the fake treatment as the independent variable. The coefficient estimate is then saved as the first observation in the RI sample. The fake treatment is then dropped, and a new fake treatment is randomly assigned. The process is repeated $r$ times. At the end of the process, one has a RI distribution of $r$ estimated treatment effects under the null hypothesis, because the fake treatment, by construction, has no effect on the outcome. This RI distribution has an expected mean of zero but a non-zero variance that captures design-based uncertainty. To get the RI $p$-value, one simply places the original estimated treatment effect in the null RI distribution. The RI $p$-value is the fraction of the null distribution that is larger in magnitude than the estimated treatment effect. This RI $p$-value can be interpreted as the probability that a similarly sized treatment effect would have been observed under the null of the treatment doing nothing.

This general procedure for RI is similar to bootstrap methods discussed above. However, while RI $p$-values are constructed similarly to bootstrapping $p$-values, there is a key difference based on the sampling reference frame. As \cite{kerwin_randomization_2017} writes, ``bootstrapped $p$-values are about uncertainty over the specific sample of the population you drew, while randomization inference $p$-values are about the uncertainty over which units within your sample are assigned to treatment''. This is classic design-based uncertainty. In bootstrapping, the assumption is that the sample of interest is representative of the population, so in sampling with replacement, bootstrapping simulates how sampling variation affects the findings. Conversely, with RI, the assumption is that treated observations in the sample are random and as such there is some non-zero probability of a treatment-control difference in outcomes, simply based on which participants were assigned to the treatment group. This assumption corresponds to designed-based uncertainty. By reassigning treatment at random in RI, this probability of a given magnitude of treatment-control difference can be ascertained. 

While RI has historically been difficult to implement, two Stata commands, from \cite{hes_randomization_2017} and \cite{young_randcmd_2020}, now make the procedures straightforward. \cite{hes_randomization_2017} details \texttt{ritest}, which conducts RI for experiments or other samples, as a whole. \cite{mckenzie_finally_2017} provides a primer on its use, which is helpful for those who might be unfamiliar. An alternative and slightly expanded command is \texttt{randcmd}. This command generates RI $p$-values for individual treatment effects and joint Wald and Westfall-Young multiple inference tests for equations with multiple treatment effects and for the experiment as whole \citep{young_randcmd_2020}. With these straightforward and easy-to-use methods, RI should become more widely used, even if simply as a robustness check.


\section{Conclusion}

This paper emphasises the need for care in calculating standard errors and other test statistics for causal inference. Given the types of data used by applied economists today, it is unlikely that the standard standard errors produced by statistical software such as Stata are the correct ones. More likely than not, the error term in an OLS regression will be heteroskedastic and standard errors will require adjustment. With more sophisticated sampling techniques and research design, it is almost assured that residuals will be correlated within groups or serially correlated. As academic papers seek to test more hypotheses, applied economists need to adjust for the probability that a true null will be rejected simply by random chance. In all of these situations, knowing what the appropriate correction procedure is requires reflection in order to articulate the exact nature of the estimand of interest and the source of uncertainty. Fortunately, the prevalence of built-in and user-written commands now makes implementation of these procedures relatively straightforward. For each challenge to conventional inference, we have provided a number of potential corrections and the relevant Stata commands. The leading difficulty is knowing which correction applies to which situation. Our goal has been to provide an intuitive understanding of what works, where, and why.

Conventional corrections to standard errors rely on asymptotic theory, which requires a variety of assumptions, such as that the data are a sample drawn from a much larger population. For many data used by applied economists today, these assumptions may not be valid. Even if the assumptions hold, numerical sampling methods, such as the bootstrap, can provide at least as accurate, and often more accurate, approximation of the distribution of the estimand or test statistics. Because of this, and the low cost of computing power, we advocate that applied economists rely on bootstrapped test statistics, not on those derived from asymptotic theory. While it is common to present bootstrapped results as a robustness check, we believe that inference relying on the bootstrap should be presented as the main result, with inference relying on asymptotic approximations as a robustness check. Additionally, when using the bootstrap, many more replications need to be done than is commonly seen in applied economics papers. To facilitate the use of the bootstrap, we have presented a variety of sampling procedures designed to address and refine asymptotic-based corrections. Again, we have provided references to built-in and user-written Stata packages to allow applied economists to easily implement these procedures. As with conventional inference, the challenge is to know how to match the procedure to the application.

Our objective has been to provide practical guidance for applied economists in how to handle the challenges to statistical inference. The tools are out there and ever growing. Careful articulation of where uncertainty comes from in the econometric model will facilitate clear thinking about which of the many methods are appropriate for the research question at hand.


\newpage
\singlespace
\bibliographystyle{chicago} 
\bibliography{inference}

\end{document}